\documentstyle[eqsecnum,aps]{revtex}
\begin{document}
\title{Double giant resonances in deformed nuclei}
\author{V.Yu. Ponomarev$^{1,2}$\thanks{%
e-mail: vlad@thsun1.jinr.ru}, C.A. Bertulani$^{1}$\thanks{%
e-mail: bertu@if.ufrj.br}, and A.V. Sushkov$^{2}$\thanks{%
e-mail: sushkov@thsun1.jinr.ru}}
\address{$^1$Instituto de F\'\i sica,
Universidade Federal do Rio de Janeiro,
21945-970 Rio de Janeiro, RJ, Brazil}
\address{$^2$Bogoliubov Laboratory of Theoretical
Physics, Joint Institute for Nuclear Research, 141980, Dubna, Russia}
\date{\today}
\maketitle

\begin{abstract}
We report on the first microscopic study of the properties of two-phonon
giant resonances in deformed nuclei. The cross sections of the excitation of
the giant dipole and the double giant dipole resonances in relativistic
heavy ion collisions are calculated. We predict that the double giant dipole
resonance has a one-bump structure with a centroid 0.8~MeV higher than twice
energy for the single giant dipole resonance in the reaction under
consideration. The width of the double resonance equals to 1.33 of that for
the single resonance.
\end{abstract}

\bigskip


\bigskip

\twocolumn

One of the most exciting progresses in the field of giant resonances in
atomic nuclei for the last few years was the experimental observation of
two-phonon giant resonances \cite{Eml94}. Nowadays, we may speak about some
systematics of their properties (the energy position, width and excitation
probability) in spherical nuclei although it is still sparse and some open
questions in this field stimulates theoretical studies (see, e.g. Refs.~\cite
{Ber93,Pon94,Pon96,Ber96a,Ber96b,Lan97,Bor97,Ber97,Pon97}). Investigation on
the properties of two-phonon giant resonances together with similar studies
on low-lying two-phonon states \cite{Kne96} should give an answer on how far
the harmonic picture of boson-type excitations holds in the finite fermion
systems like atomic nuclei.

The possibility to observe two-phonon giant resonances in deformed nuclei
with the present state of art experimental techniques is still questionable.
This is mainly due to the fact that one has to expect a larger width of
these resonances as compared to spherical nuclei. Also, the situation with
the low-lying two-phonon states in deformed nuclei is much less clear than
in spherical ones.

The first experiment with the aim to observe the double giant dipole
resonance (DGDR) in $^{238}$U in relativistic heavy ion collisions (RHIC)
was performed recently at the GSI/SIS facility by the LAND collaboration 
\cite{Eml-pr}. It will take some time to analyze the experimental data and
to present the first experimental evidence of the DGDR in deformed nuclei,
if any. The first microscopic study of properties of the DGDR in deformed
nuclei is the subject of the present paper. The main attention will be paid
to the width of the DGDR and its shape.

In a phenomenological approach the GDR is considered as a collective
vibration of protons against neutrons. In spherical nuclei this state is
degenerate in energy for different values of the spin $J=1^{-}$ projection $%
M=0,\pm 1$. The same is true for the $2^{+}$ component of the DGDR with
projection $M=0,\pm 1,\pm 2$. In deformed nuclei with an axial symmetry like 
$^{238}$U, the GDR is spit into two components $I^\pi (K)=1^{-}(0)$ and $%
I^\pi (K)=1^{-}(\pm 1)$ corresponding to vibrations against two different
axes. In this approach one expects a three-bump structure for the DGDR with
the value $K=0$, $K=\pm 1$ and $K=0,\pm 2$, respectively. Actually, the GDR
possesses a width and the main mechanism responsible for it in deformed
nuclei is the Landau damping. Thus, the conclusion on how three bumps
overlap and what is the real shape of the DGDR in these nuclei, i.e., either
a three-bump or a flat broad structure, can be drawn out only from some
consistent microscopic studies.

In the present paper we use the Quasiparticle Phonon Model (QPM) \cite{Sol92}
to investigate the properties of the GDR and the DGDR in $^{238}$U. The QPM,
although with somewhat different technical details which reflect the
difference between spherical and deformed nuclei, was used to investigate
the same resonances in spherical nuclei in Refs.~\cite
{Pon94,Pon96,Ber96b,Pon97}). The model Hamiltonian includes an average field
for protons and neutrons, monopole pairing and residual interaction in a
separable form. We use in our calculations for $^{238}$U the parameters of
Woods-Saxon potential for the average field and monopole pairing from the
previous studies \cite{param}. They were adjusted to reproduce the
properties of the ground state and low-lying excited states. The average
field has a static deformation with the deformation parameters $\beta _2=0.22
$ and $\beta _4=0.08$. To construct the phonon basis for the $K=0$ and $%
K=\pm 1$ components of the GDR we use the dipole-dipole residual interaction
(for more details, see e.g. Ref.~\cite{Sol92}). The strength parameters of
this interaction are taken from Ref.~\cite{Mal} where they have been fitted
to obtain the centroid of the B(E1, $0_{g.s.}^{+}\rightarrow 1^{-}(K=0,\pm
1))$ strength distribution at the value known from experiment \cite{Gur76}
and to exclude the centre of mass motion. In this approach, the information
on the phonon basis (i.e. the excitation energies of phonons and their
internal fermion structure) is obtained by solving the RPA equations. For
electromagnetic E1-transitions we use the free values of the effective
charges, $e_{eff}^{Z(N)}=eN(-Z)/A$.

The results of our calculation of the B(E1) strength distribution over $%
\left| 1_{K=0}^{-}(i)\right\rangle $ and $\left| 1_{K=\pm 1}^{-}(i^{\prime
})\right\rangle $ GDR states are presented in Fig.~\ref{fig1}, together with
the experimental data. The index $i$ in the wave function stands for the
different RPA states. All one-phonon states with the energy lower than
20~MeV and with the B(E1) value larger than $10^{-4}\ e^2fm^2$ are accounted
for. Their total number equals to 447 and 835 for the $K=0$ and $K=\pm 1$
components, respectively. Only the strongest of them with B(E1)$\ge 0.2\
e^2fm^2$ are shown in the figure by vertical lines. Our phonon basis
exhausts 32.6\% and 76.3\% of the energy weighted sum rules, $14.8\cdot NZ/A$
$e^2$ fm$^2$ MeV, by the $K=0$ and $K=\pm 1$ components, respectively. For a
better visual appearance we also present in the same figure the strength
functions averaged with a smearing parameter, which we take as 1~MeV. The
long (short) dashed-curve represent the $K=0$ ($K=\pm 1)$ components of the
GDR. The solid curve is their sum. The calculation reproduces well the
two-bump structure of the GDR and the larger width of its $K=\pm 1$
component. The last is consistent with the experiment \cite{Gur76} which is
best fitted by two Lorentzians with widths equal to $\Gamma _1=2.99$~MeV and 
$\Gamma _2=5.10$~MeV, respectively. The amplitudes of both maxima in the
calculation are somewhat overestimated as compared to the experimental data.
This happens because the coupling of one-phonon states to complex
configurations is not taken into account which can be more relevant for the $%
K=\pm 1$ peak at higher energies. But in general the coupling matrix
elements are much weaker in deformed nuclei as compared to spherical ones
and the Landau damping describes the GDR width on a reasonable level.

The wave function of the $0^{+}$ and $2^{+}$ states belonging to the DGDR
are constructed by the folding of two $1^{-}$ phonons from the previous
calculation. When a two-phonon state is constructed as the product of two
identical phonons its wave function gets an additional factor $1/\sqrt{2}$.
The $1^{+}$ component of the DGDR is not considered here since its
excitation is quenched in RHIC for the same reasons as in spherical nuclei 
\cite{Ber96b}. The anharmonicity effects which arise from interactions
between different two-phonon states are also not included in the present
study. It was shown that these effects have an $A^{-4/3}$ dependence on the
mass number $A$ \cite{Ber97} and that they are small for the DGDR \cite
{Pon96,Lan97,Ber97} even for $^{136}$Xe and $^{208}$Pb.

The folding procedure yields three groups of the DGDR states: \\ 
\begin{eqnarray}
&&a)~~\left| [1_{K=0}^{-}(i_1)\otimes
1_{K=0}^{-}(i_2)]_{0_{K=0}^{+},2_{K=0}^{+}}\right\rangle ,  \nonumber \\
&&b)~~\left| [1_{K=0}^{-}(i)\otimes 1_{K=\pm 1}^{-}(i^{\prime })]_{2_{K=\pm
1}^{+}}\right\rangle ~~\mbox{and}  \nonumber \\
&&c)~~\left| [1_{K=\pm 1}^{-}(i_1^{\prime })\otimes 1_{K=\pm
1}^{-}(i_2^{\prime })]_{0_{K=0}^{+},2_{K=0,\pm 2}^{+}}\right\rangle .
\label{wf}
\end{eqnarray}
The total number of non-degenerate two-phonon states equals to about $%
1.5\cdot 10^6$. The energy centroid of the first group is the lowest and of
the last group is the highest among them. So, we also obtain the three-bump
structure of the DGDR. But the total strength of each bump is fragmented
over a wide energy region and they strongly overlap.

Making use of the nuclear structure elements discussed above, we have
calculated the excitation of the DGDR in $^{238}$U projectiles (0.5~GeV$%
\cdot $A) incident on $^{120}$Sn and $^{208}$Pb targets, following the
conditions of the experiment in Ref.~\cite{Eml-pr}. These calculations have
been performed in the second order perturbation theory \cite{Ber88}, in
which the DGDR states of Eq.~(\ref{wf}) are excited within a two-step
process: g.s.$\rightarrow $GDR$\rightarrow $DGDR. As intermediate states,
the full set of one-phonon $\left| 1_{K=0}^{-}(i)\right\rangle $ and $\left|
1_{K=\pm 1}^{-}(i^{\prime })\right\rangle $ states was used. We have also
calculated the GDR excitation to first order for the same systems. The
minimal value of the impact parameter, which is very essential for the
absolute values of excitation cross section has been taken according to $%
b_{min}=1.28\cdot (A_t^{1/3}+A_p^{1/3})$.

The results of our calculations are summarized in Fig.~\ref{fig2} and Table~%
\ref{tab1}. In Fig.~\ref{fig2} we present the cross sections of the GDR
(part a) and the DGDR (part b) excitation in the $^{238}$U (0.5~GeV$\cdot $%
A) + $^{208}$Pb reaction. We plot only the smeared strength functions of the
energy distributions because the number of two-phonon states involved is
numerous. The results for $^{238}$U (0.5~GeV$\cdot $A) + $^{120}$Sn reaction
look very similar and differ only by the absolute value of cross sections.
In Table~\ref{tab1} the properties of the GDR and the DGDR, and their
different $K$ components are given. The energy centroid $E_c$ and the second
moment, $m_2=\sqrt{\sum_k\sigma _k\cdot (E_k-E_c)^2/\sum_k\sigma _k},$ of
the distributions are averaged values for the two reactions under
consideration.

The two-bump structure can still be seen in the curve representing the cross
section of the GDR excitation in $^{238}$U in RHIC as a function of the
excitation energy. But its shape differs appreciably from the B(E1) strength
distribution (see Fig.~\ref{fig2}a in comparison with Fig.~\ref{fig1}). The
reason for that is the role of the virtual photon spectra. First, for the
given value of the excitation energy and impact parameter it is larger for
the $K=\pm 1$ component than that for the $K=0$ one (see also the first two
lines in Table~\ref{tab1}). Second, for both components it has a decreasing
tendency with an increase of the excitation energy \cite{Ber88}. As a
result, the energy centroid of the GDR excitation in RHIC shifts by the
value 0.7~MeV to lower energies as compared to the same value for the B(E1)
strength distribution. The second moment $m_2$ increases by 0.2~MeV.

The curves representing the cross sections of the excitation of the $K=\pm 1$
and $K=\pm 2$ components of the DGDR in $^{238}$U in RHIC have typically a
one-bump structure (see the curves with squares and triangles in Fig.~\ref
{fig2}b, respectively). It is because they are made of two-phonon $2^{+}$
states of one type: the states of Eq.~(\ref{wf}b) and Eq.~(\ref{wf}c),
respectively. Their centroids should be separated by an energy approximately
equal to the difference between the energy centroids of the $K=0$ and $K=\pm
1$ components of the GDR. They correspond to the second and the third bumps
in a phenomenological treatment of the DGDR. The $K=0$ components of the
DGDR include two group of states: the states represented by Eq.~(\ref{wf}a)
and those of Eq.~(\ref{wf}c). Its strength distribution has two-bumps (see
the curve with circles for the $2^{+}(K=0)$ and the dashed curve for the $%
0^{+}(K=0)$ components of the DGDR, respectively). The excitation of the
states given by Eq.~(\ref{wf}a) in RHIC is enhanced due to their lower
energies, while the enhancement of the excitation of the states given by
Eq.~(\ref{wf}c) is related to the strongest response of the $K=\pm 1$
components to the external E1 Coulomb field in both stages of the two-step
process.

Summing together all components of the DGDR yields a broad one-bump
distribution for the cross section for the excitation of the DGDR in $^{238}$%
U, as a function of excitation energy. It is presented by the solid curve in
Fig.~\ref{fig2}b. Another interesting result of our calculations is related
to the position of the DGDR energy centroid and to the second moment of the
DGDR cross section. The centroid of the DGDR in RHIC is shifted to the
higher energies by about 0.8~MeV from the expected value of two times the
energy of the GDR centroid. The origin for this shift is in the energy
dependence of the virtual photon spectra and it has nothing to do with
anharmonicities of the two-phonon DGDR states. In fact, the energy centroid
of the B(E1, $g.s.\rightarrow 1_i^{-})\times $ B(E1, $1_i^{-}\rightarrow $%
DGDR$_f)$ strength function appears exactly at twice the energy of the
centroid of the B(E1, $g.s.\rightarrow $GDR) strength distribution because
the coupling between different two-phonon DGDR states are not accounted for
in the present calculation. The same shift of the DGDR from twice the energy
position of the GDR in RHIC also takes place in spherical nuclei. But the
value of the shift is smaller there because in spherical nuclei the GDR and
the DGDR strength is less fragmented over their doorway states due to the
Landau damping. For example, this shift equals to 0.25~MeV in $^{208}$Pb for
the similar reaction. This effect is also seen when the DGDR position
against the GDR is reported from experimental studies \cite{Bor96}. But the
larger value of the shift under consideration in deformed nuclei should
somehow simplify the separation of the DGDR from the total cross section in
RHIC.

Another effect which also works in favor of the extraction of the DGDR from
RHIC excitation studies with deformed nuclei is its smaller width than $%
\sqrt{2}$ times the width of the GDR, as observed with spherical nuclei. Our
calculation yields the value 1.33 for the ratio $\Gamma {\mbox {\tiny DGDR}}%
/\Gamma {\mbox {\tiny GDR}}$ in this reaction. The origin for this effect is
in the different contributions of the GDR $K=0$ and $K=\pm 1$ components to
the total cross section, due to the reaction mechanism. It should be
remembered that only the Landau damping is accounted for the width of both
the GDR and the DGDR. But we think that the effect of narrowing of the DGDR
width still holds if the coupling to complex configurations is included in
the calculation.

To conclude, we present the first theoretical studies based on microscopic
calculation of the properties of the two-phonon giant dipole resonance in
deformed nuclei in relativistic heavy ion collisions. We predict that the
excitation function has a one-bump shape and that there are at least two
effects which work in favor of its experimental observation, namely, the
energy shift to higher energies, and the narrowing of its width.

V.Yu. P. thanks the Instituto de F\'{\i}sica of the Universidade Federal do
Rio de Janeiro for the hospitality, and the CNPq for financial support. We
thank Dr. L.A. Malov for fruitful discussions. This work was partially
supported by the RFBR (grant 96-15-96729), by the FUJB/UFRJ, and by the
MCT/FINEP/CNPq(PRONEX) (contract 41.96.0886.00)

\begin{table}[tbh]
\caption{The properties of the different components of the GDR and the DGDR
in $^{238}$U. The energy centroid $E_c$, the second moment of the strength
distribution $m_2$ in RHIC, and the cross sections $\sigma $ for the
excitation of the projectile are presented for: a) $^{238}$U (0.5~GeV$\cdot $%
A) + $^{120}$Sn, and b) $^{238}$U (0.5~GeV$\cdot $A) + $^{208}$Pb.}
\label{tab1}
\begin{tabular}{lccrr}
& $E_{c}$ & $m_2$ & \multicolumn{2}{c}{$\sigma$ [mb]} \\ 
& [MeV] & [MeV] & a)\hspace*{2mm} & b)\hspace*{2mm} \\ \hline
GDR($K=0$) & 11.0 & 2.1 & 431.2 & 1035.4 \\ 
GDR($K=\pm1$) & 12.3 & 2.6 & 1560.2 & 3579.1 \\ 
GDR(total) & 12.0 & 2.6 & 1991.4 & 4614.5 \\ \hline
DGDR$_{0^+}$($K=0$) & 25.0 & 3.4 & 18.3 & 88.9 \\ 
DGDR$_{2^+}$($K=0$) & 24.4 & 3.5 & 11.8 & 58.7 \\ 
DGDR$_{2^+}$($K=\pm1$) & 23.9 & 3.2 & 22.7 & 115.4 \\ 
DGDR$_{2^+}$($K=\pm2$) & 25.3 & 3.4 & 49.7 & 238.3 \\ 
DGDR(total) & 24.8 & 3.4 & 102.5 & 501.3
\end{tabular}
\end{table}

\begin{figure}[tbh]
\caption{The B(E1) strength distribution over $K=0$ (short-dashed curve) and 
$K=\pm 1$ (long-dashed curve) $1^{-}$ states in $^{238}$U. The solid curve
is their sum. The strongest one-phonon $1^{-}$ states are shown by vertical
lines, the ones with $K=0$ are marked by a triangle on top. Experimental
data are from Ref.~{\protect\cite{Gur76}}.}
\label{fig1}
\end{figure}

\begin{figure}[tbh]
\caption{The strength functions for the excitation: a) of the GDR, and b) of
the DGDR in $^{238}$U in the $^{238}$U (0.5~GeV$\cdot $A) + $^{208}$Pb
reaction. In a), the short-dashed curve corresponds to the GDR ($K=0$) and
the long-dashed curve to the GDR ($K=\pm 1$). In b) the dashed curve
corresponds to the DGDR$_{0^{+}}$ ($K=0$), the curve with circles to the DGDR%
$_{2^{+}}$($K=0$), the curve with squares to the DGDR$_{2^{+}}$ ($K=\pm 1$),
and the curve with triangles to the DGDR$_{2^{+}}$ ($K=\pm 2$) . The solid
curve is the sum of all components. The strength functions are calculated
with the smearing parameter equal to 1 MeV.}
\label{fig2}
\end{figure}

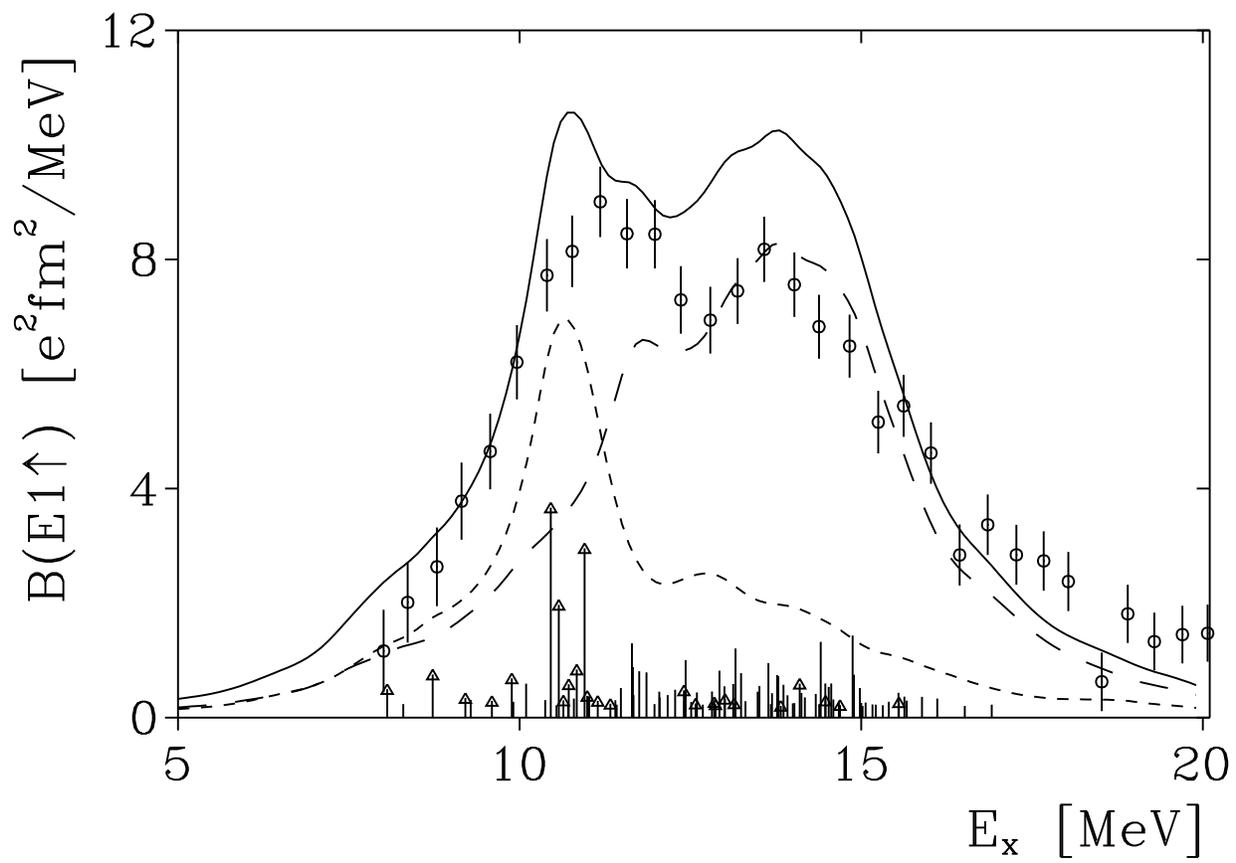
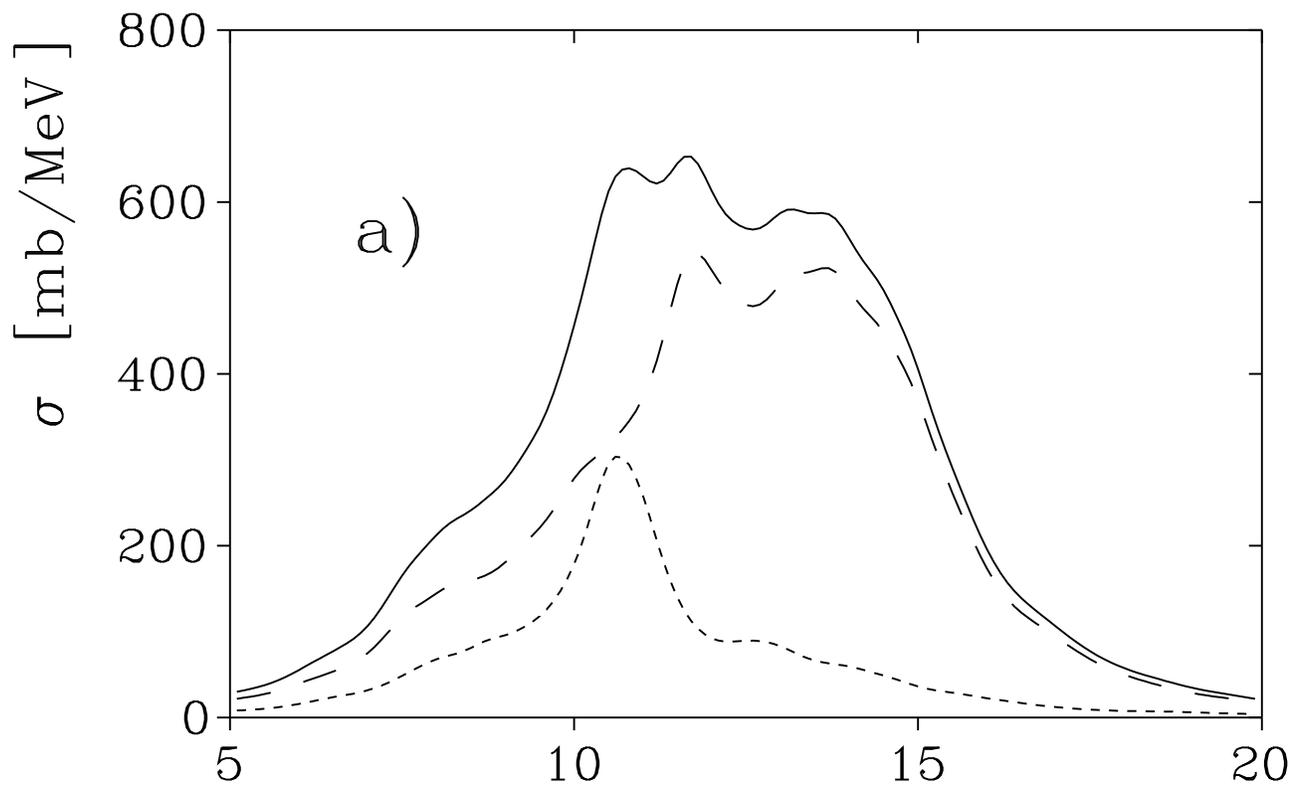
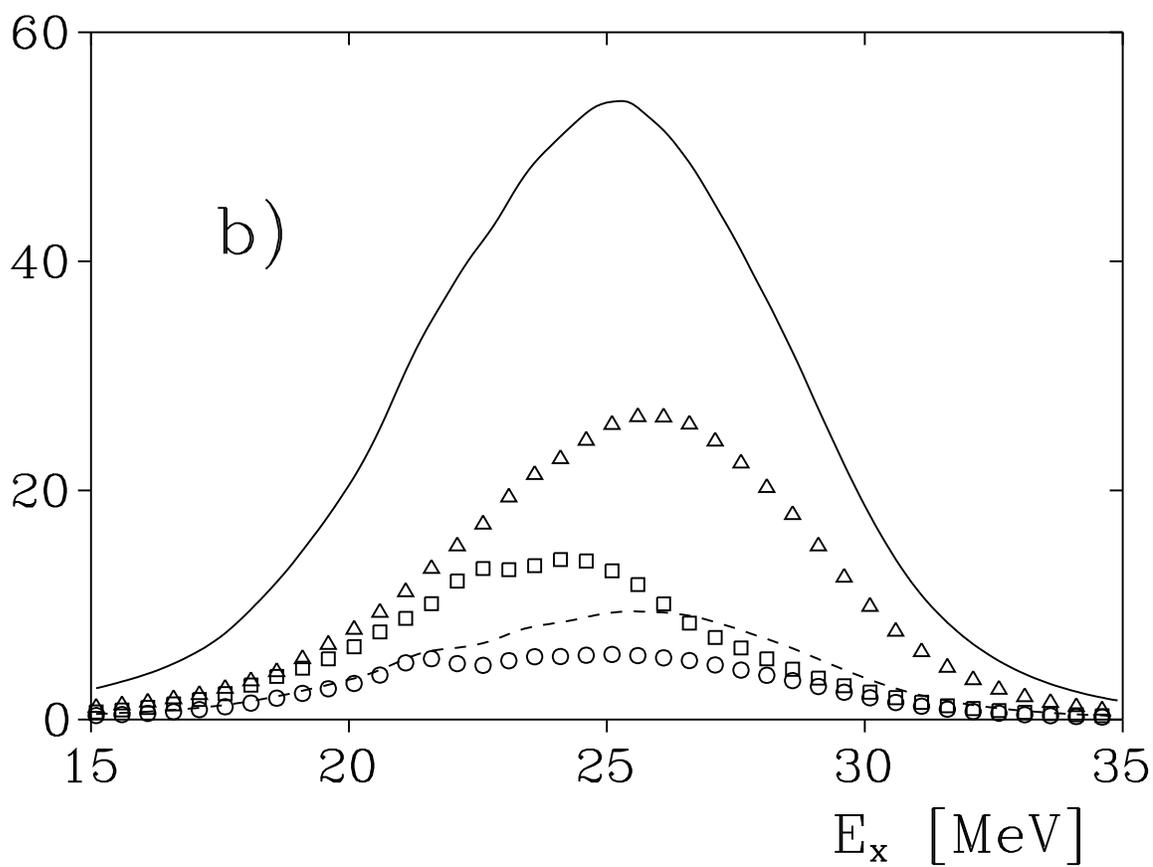
\end{document}